\documentclass[12pt]{article}
\title{The matrix Hamiltonian for hadrons  and the role of negative-energy components}
\author{  Yu.A.Simonov\\
 State Research Center\\Institute of Theoretical and Experimental
Physics, \\ Moscow, 117218 Russia}
 \date{}
\newcommand{\beq}{\begin{eqnarray}}
 \newcommand{\eeq}{\end{eqnarray}}
\newcommand{\be}{\begin{equation}}
 \newcommand{\ee}{\end{equation}}

 \def\la{\mathrel{\mathpalette\fun <}}

\def\fun#1#2{\lower3.6pt\vbox{\baselineskip0pt\lineskip.9pt
\ialign{$\mathsurround=0pt#1\hfil ##\hfil$\crcr#2\crcr\sim\crcr}}}

\newcommand{{\SD}}{\rm SD}

\newcommand{\vex}{\mbox{\boldmath${\rm x}$}}
\newcommand{\vey}{\mbox{\boldmath${\rm y}$}}
\newcommand{\ver}{\mbox{\boldmath${\rm r}$}}
\newcommand{\vesig}{\mbox{\boldmath${\rm \sigma}$}}

\newcommand{\vep}{\mbox{\boldmath${\rm p}$}}

\newcommand{\vez}{\mbox{\boldmath${\rm z}$}}

\newcommand{\ves}{\mbox{\boldmath${\rm s}$}}

\newcommand{\ven}{\mbox{\boldmath${\rm n}$}}
\newcommand{\veu}{\mbox{\boldmath${\rm u}$}}

\newcommand{\veB}{\mbox{\boldmath${\rm B}$}}

\newcommand{\veE}{\mbox{\boldmath${\rm E}$}}

\newcommand{\veal}{\mbox{\boldmath${\rm \alpha}$}}

\newcommand{\llan}{\langle\langle}
\newcommand{\rran}{\rangle\rangle}
\newcommand{\lan}{\langle}
\newcommand{\ran}{\rangle}
\begin{document}
\maketitle

\begin{abstract}
The world-line (Fock-Feynman-Schwinger) representation is used for
quarks in arbitrary (vacuum and valence gluon) field to construct
the relativistic Hamiltonian. After averaging the Green's function
of the white $q\bar q$  system over gluon fields  one obtains the
relativistic Hamiltonian, which is matrix in spin indices and
contains both positive and negative quark energies.

The role of the latter is  studied in the example of the
heavy-light meson and the standard einbein  technic  is extended
to the case of the matrix Hamiltonian. Comparison with the Dirac
equation shows a good agreement of the results. For arbitrary
$q\bar q $ system the nondiagonal matrix Hamiltonian components
are calculated through hyperfine interaction terms. A general
discussion of the role of negative energy components is given in
conclusion.

\end{abstract}

\section{ Introduction}

The quest for the Hamiltonian which contains main features of QCD
- confinement and Chiral Symmetry Breaking (CSB) exists from the
very beginning, when fundamental field-theoretical QCD
Hamiltonians, have been constructed in different gauges \cite{1}.
Unfortunately (nonlocal) confinement cannot be seen in these local
FTh Hamiltonians and for practical  purposes another sort of
Hamiltonians --  Effective Hamiltonians  (EH) have been modelled
containing minimal relativity and string-type potentials \cite{2}.
A lot of  information was  obtained from these Hamiltonians and
the general  agreement of calculated meson and baryon masses with
experiment is impressive \cite{3}, with some exceptions for mesons
(e.g. pions, scalar nonets etc.) and for baryons (the Roper
resonance  and his companions, $\Lambda(1405) $ etc.).

The two main defects of effective Hamiltonians are that: i) The
clear-cut derivation  from the basic QCD Lagrangian was absent and
therefore it is not clear what are approximations and how to
improve EH systematically.

ii) Connected to that, the EH contains a large number of
parameters  in addition to the minimal QCD number: current quark
masses and string tension (or $\Lambda_{QCD}$). Typically this
additional number of parameters is more than ten for detailed
spectrum calculation. The most important for hadron masses are
constituent quark masses, $m_i$ and the overall negative constant
$C_0$ of the order of several hundred MeV.

With this number of arbitrary parameters the QCD dynamics in
hadrons cannot be fully understood and one needs another approach.
This approach (it will be called the QCD string approach (SA)) was
suggested more than a decade ago  \cite{4}, where the SA
Hamiltonian for spinless quarks was derived and Regge trajectories
have been obtained both for mesons \cite{4} baryons \cite{5}, and
for glueballs \cite{6}. Later on, the formalism was put on more
rigorous basis in \cite{7} and the einbein  technic \cite{8} was
used in \cite{7} to take into account the string moment of inertia
and to obtain the correct Regge slope. In \cite{4}-\cite{7} the
constituent mass was defined using the einbein technic through the
string tension and current quark mass; the subsequent calculation
of baryon magnetic moments \cite{9} has confirmed the validity of
this approach.

Another mysterious problem -- of large negative constant $C_0$ --
 was understood recently in the framework of the same QCD string
 approach and $C_0$ was identified with the large quark
 self-energy term \cite{10}. The latter is calculated through the
 string  tension and quark current masses again without
 introduction of new parameters.

 The Spin-Dependent (SD) part of the SA Hamiltonian was calculated
 earlier \cite{11}. It was shown that even for light quarks one
 can calculate the leading SD terms without recurring to the $1/M$
 expansion, but using instead the lowest (quadratic)
 field-correlator approximation \cite{12} which works with
 accuracy of few percents \cite{13}.

The final form of the SA Hamiltonian was used to calculate the
masses of light mesons \cite{14,15}, heavy quarkonia
\cite{16,17,18}, heavy-light mesons \cite{19,20,21}, baryons
\cite{22,23,24}, glueballs \cite{6,25,26},
hybrids\cite{27,28,29,30}, gluelumps\cite{31}. For a review see
also \cite{32,33,34}. It is remarkable that in most cases the
agreement with known experimental data and lattice data is good,
however only the minimal QCD  set of parameters was used with
addition of standard $\alpha_s(Q^2).$

The most important exception in mesons from the agreement above
was for pseudoscalars ($\pi, K, \eta, \eta'$) which need the
chiral dynamics absent in the SA Hamiltonian. To overcome this
discrepancy it was realized in \cite{35,36,37} that the chiral
dynamics  brings  a new tadpole term, which should be accounted
for  in computation of the Nambu-Goldstone spectrum. As a result
the Gell-Mann-Oakes-Renner relation was found in \cite{35,36} and
the quark condensate was computed \cite{37} in terms of the SA
Hamiltonian spectrum.

This connection allows to calculate the spectrum of
Nambu-Goldstone mesons and their radial excitations in terms of
the SA Hamiltonian spectrum, without introducing new  parameters.

So far so good, but to proceed further one should look carefully
into the approximations done and understand how to improve the SA
Hamiltonian systematically.

The systematic procedure of the derivation of the SA Hamiltonian
is given in \cite{4}-\cite{7} and  discussed later in
\cite{38}-\cite{40}. It contains three typical approximations: 1)
Replacement of the Wilson loop by the minimal area expression and
neglect of gluon excitation, which amounts to the neglect of
mixing of a given hadron with all its hybrid excitations. As it
was shown in \cite{41} the effect of mixing is indeed small except
for the cases of states almost degenerate in mass. 2) The use of
the local Hamiltonian which appears in the limit of small gluon
correlation length $\lambda$ (denoted as $T_g$ in most previous
publications). Since $\lambda\cong 0.2$ fm and much smaller than
typical hadron sizes, this limit is legitimate. 3) The use of only
positive solution for the stationary point equations in the
einbein variable, corresponding to the  positive constituent quark
mass. This latter approximation means neglect of the quark
negative energy states, and it is the main point of the present
investigation. As a result we shall obtain the Hamiltonian
containing both positive and negative quark components, and
estimate quantitatively the importance of the latter.

 The paper is organized as follows. In the section 2 we derive the
 Green function for the $q\bar q$ system and consider in section 3
      the one-body self-energy corrections for
      the quark and antiquark.
       Having fixed that, we turn
       in section 4 to the  heavy-light $q\bar q$ interaction and derive the
       corresponding $q\bar q$ Hamiltonian
   in the full relativistic form,
    containing Negative Energy Components (NEC) and compare numerical results of matrix Hamiltonian with
   those for the Dirac equation.
    In the section 5  the effects of NEC are derived for the general $q\bar q$ system.
     The last section is devoted to conclusions and outlook.
 Appendix 1 is devoted to the derivation
    of path integral form of the FFS type, in particular
     a novel first-order form of FFS is obtained for
     one particle in external nonabelian field using Weyl representation for $\gamma$ matrices.
      Appendix 2 contains details of self-energy correction.

\section{The  quark-antiquark Green function}

We  recapitulate here  the steps done in derivation of the SA
Hamiltonian \cite{4,7,32}.

One starts with the Fock-Feynman-Schwinger Representation (FFSR)
for  the quark (or valence gluon) Green's function in  the
Euclidean external gluonic fields \cite{34,41}, which is exact and
does not contain any approximation:
\be
S(x,y) = (m+\hat D)^{-1} = (m-\hat D) \int^\infty_0 ds (Dz)_{xy}
e^{-K}P_A\exp (ig \int^x_y A_\mu dz_\mu) P_\sigma (x, y,s)
\label{1}\ee where $K$ is the kinetic energy term,
\be
K= m^2s+ \frac14 \int^s_0 d\tau \left(\frac{dz_\mu(\tau)}{d\tau}
\right)^2\label{2}\ee and $m$ is the pole mass of quark, and
$z_\mu(\tau)-$ the quark trajectory with end points $x$ and $y$
integrated over in $(D z)_{xy}$.

The factor $P_\sigma (x,y,s)$ in (\ref{1}) is generated by the
quark spin (color-magnetic moment) and is equal
\be
P_\sigma(x,y,s)= P_F\exp [g\int^s_0 \sigma_{\mu\nu} F_{\mu\nu}
(z(\tau))d\tau],\label{3}\ee where $\sigma_{\mu\nu} =\frac{1}{4i}
(\gamma_\mu\gamma_\nu-\gamma_\nu\gamma_\mu),$ and $P_F(P_A)$ in
(\ref{3}), (\ref{1}) are respectively  ordering operators of
matrices $F_{\mu\nu} (A_\mu)$ along the path $z_\mu(\tau)$. In
what follows the role of the operator $P_\sigma(x,y,s)$ will be
crucial, and it is  convenient to rewrite $\sigma_{\mu\nu}
F_{\mu\nu}$ in $2\times 2$ notations
\be
\sigma_{\mu\nu} F_{\mu\nu} = \left(\begin{array}{ll} \vesig \veB&
\vesig\veE\\ \vesig \veE& \vesig \veB\end{array}\right)
\label{4}\ee where $\vesig$ are usual Pauli matrices.

The next step  is the FFSR  for the hadron Green's function, which
for the case of the $q \bar q$ meson is
\be
 G_{q\bar q}(x,y;A) =
\int^\infty_0 ds \int^\infty_0 ds'(Dz)_{xy} (Dz')_{xy} e^{-K-K'}
tr (\Gamma(m-\hat D) W_{\sigma\sigma} (x,y) \bar \Gamma (m'-\hat
D'))\label{5}\ee where $\Gamma$ and $ \bar \Gamma= \Gamma^+$ are
1, $\gamma_\mu,\gamma_5, (\gamma_\mu\gamma_5),...$, "tr" means
trace operation both in Dirac and color indices, and
\be
W_{\sigma\sigma'} (x,y) = P_A \exp (ig \int_{C(x,y)} A_\mu d
z_\mu) P_\sigma (x,y,s) P'_\sigma (x,y, s').\label{6}\ee In
(\ref{6}) the closed contour $C(x,y)$ is along trajectories of
quark $z_\mu(\tau)$ and antiquark  $z'_\nu(\tau')$, and the
ordering $P_A$ and $P_F$ in  $P_\sigma,P'_\sigma$, is universal,
i.e. $W_{\sigma \sigma'}  (x,y)$  is the Wegner-Wilson loop with
insertion of operators  (\ref{4}) along the contour $C(x,y)$ at
the proper places.

The FFSR Eq. (\ref{5}) is exact and is a functional of gluonic
fields $A_\mu, F_{\mu\nu}$, which  contain both perturbative and
nonperturbative contributions, not specified at this level.

The next step, containing important approximation, is the
averaging over gluonic fields, which yields the physical $q \bar
q$ Green's function $G_{q\bar q}$ \be G_{q\bar q} (x,y) =\lan
G_{q\bar q} (x,y;A)\ran_A.\label{7}\ee

Here the averaging is done with the usual Euclidean weight $\exp
(-action)$, containing all gauge-fixing, ghost terms, the exact
form is inessential for what follows. To proceed it is convenient
to use the nonabelian Stokes theorem \cite{42} for the first
factor on the r.h.s. in (\ref{6}) and to rewrite the average of
(\ref{6}) as$$ \lan W_{\sigma\sigma'} (x,y) \ran = \lan \exp ig
\int d\pi_{\mu\nu} (z) F_{\mu\nu} (z)\ran=$$ \be= \exp
\sum^\infty_{n=1} \frac{(ig)^n}{n!} \int d\pi(1)... \int d\pi (n)
\llan F(1)...F(n)\rran\label{8}\ee where \be d\pi_{\mu\nu} (z)= d
s_{\mu\nu} (z) -i \sigma_{\mu\nu} d\tau\label{9}\ee and
$ds_{\mu\nu}$ is the surface element. In (\ref{8}) we have used
the cluster expansion theorem  and omitted indices of $d\pi(k),
F(k)$ implying $d\pi(k) \equiv d\pi_{\mu_k\nu_k}(z) =
ds_{\mu\nu}(u_k)-i\sigma_{\mu_k\nu_k}d\tau_k, F(k) \equiv
F_{\mu_k\nu_k} (u_k, x_0) \equiv \Phi(x_0, u_k) F_{\mu_k\nu_k}
(_k) \Phi(u_k, x_0),$ where $\Phi(x,y)= P\exp ig \int^x_y A_\mu
dz_\mu$.

Eq. (\ref{8}) is exact and therefore the r.h.s. does not depend on
the choice of the surface, which is integrated over in
$ds_{\mu\nu} (z)$. To proceed one makes at this point  the
approximation, keeping only lowest (quadratic) field correlator
$\llan F(i) F(k)\rran$, while the surface is chosen to be the
minimal area surface. As it was argued in \cite{13}, using
comparison with lattice data, this approximation (sometimes called
the Gaussian Approximation (GA)) has accuracy of few percent. The
factors $(m-\hat D)$ and $(m'-\hat D')$ in (\ref{5}) need  a
special treatment in the process of averaging in (\ref{7}), and as
shown in Appendix 1 of \cite{19} one can use a simple replacement,
\be
 m-\hat D\to  m -i \hat p, ~~ p_\mu =\frac{1}{2}
\left ( \frac{dz_\mu}{d\tau}\right)_{\tau=s}.\label{10} \ee

 With the insertion of the cluster expansion (\ref{8}) and the
 operator $(m-\hat D)$ from  Eq.(\ref{10}) into the general
 expressions (\ref{7}), (\ref{5}), one fulfills the first step:
 the derivation of the physical $q\bar q$ Green's function in
 terms of vacuum correlators $\llan F(1)...F(n)\rran$. At this
 point it is important to discuss the separation of one-body
 (self-energy) and two-body terms in the interaction kernel
 (\ref{8}), together with the separation of perturbative and
 nonperturbative contributions.

 \section{Quark self-energy in the confining background}

It is clear that on physical grounds it is difficult to separate
out the one-body (self-energy) contributions for the quark
connected by the string to the antiquark. The situation here is
different in the confining QCD from the nonconfining QED, since in
the latter electron can be isolated and its selfenergy is a part
of the renormalized electron mass operator. For the bound electron
in an atom, the one body and two-body contributions to the
interaction kernel can be separated in each order in $\alpha^n$,
as it is done e.g. in the Bethe-Salpeter equation.

In case of QCD confinement cannot be excluded in any order of
perturbative gluon exchanges and the separation seems to be
impossible in principle. Nevertheless the world-line or FFS
representation (\ref{1}), (\ref{5}), (\ref{7}) suggests a possible
way of separating out the one-body contributions, It is based on
the  distinguishing the perimeter $(L)$ law and area  $(S_{min})$
law terms in the Wegner-Wilson loop, \be \lan W(C)\ran =const\exp
(-C_1 L-C_2 S_{min}),\label{11}\ee where the one-body terms is
associated with the coefficient $C_1, $ while the two-body terms
-- with $C_2$. Going over from the Wegner-Wilson loop to the
$q\bar q$ Green's  function and the $q\bar q$ Hamiltonian, the
situation however is becoming more complicated since $\lan
G_{q\bar q}\ran$ is an integral over all Wegner-Wilson loops, and
typical loops ($q\bar q$ trajectories) have a finite  average
$q\bar q$ separation $\lan r\ran$ and the same time length $T$, so
that both perimeter and area-law terms contribute terms
proportional to $T$. At this point  the FFSR is helpful, since it
allows to separate the Lorentz-invariant self-energy (SE) terms,
$\Delta m^2$, which contribute to the Hamiltonian (see below and
in \cite{10}) as $\frac{\Delta m^2}{2\omega}$, where $2\omega =
\frac{dt}{ds}$, and $t$ is the physical time, and $s$ , as in
(\ref{1}), is the proper time. One can see that  the SE  terms are
multiplied by the effective length of trajectory, indeed $\Delta H
dt = \frac{\Delta m^2}{2\omega} dt \sim \Delta m^2 ds$. At the
same time the two-body terms in the Hamiltonian are proportional
to $\omega$ (see below and in \cite{10,32}).

To calculate the SE terms explicitly we shall use the  background
perturbation theory \cite{43,44} with the separation of
nonperturbative background field $B_\mu$ and valence
(perturbative) gluon field $a_\mu$, so that the total vector
potential $A_\mu$ can be written as
\be
A_\mu = B_\mu+a_\mu.\label{12}\ee The method \cite{44} assumes the
perturbative expansion in powers of $ga_\mu$, while $B_\mu$ enters
via nonperturbative field correlators known from lattice \cite{45}
or analytic \cite{31} calculations.
 Accordingly we
separate the contributions to the quark SE (we prefer to use the
$m^2$ instead of $m$, since $m^2$ appears in the Hamiltonian both
in the FFS technic and after Foldy-Wouthuyzen diagonalization of
the Dirac operators)
\be
m^2(\mu) = m^2_{pert} (\mu) + \Delta m^2_{np} (\mu) +
m^2_{int}.\label{13}\ee Here $m^2_{pert}(\mu)$ is the pole mass
and its connection to the $\overline{MS}$ mass is known to two
loops (for a detailed discussion see the book \cite{46})
\be
m_{pert}^{(pole)} (\mu) = \overline{m}(\overline{m}^2)\left \{ 1+
\frac{C_F}{\pi} \alpha_s (m^{pole})
+O(\alpha^2_s)\right\},\label{14}\ee while the basic
nonperturbative term $m^2_{np}$ was found in \cite{10} (below we
shall find a correction to this term) and the mixed
perturbative-nonperturbative contribution $m^2_{int}$ is yet to be
calculated. We stress that $m^2(\mu)$ can be found in a
gauge-invariant form only when it is computed inside the
gauge-invariant $q\bar q$ or $3q$ Green's function, and in
principle it may depend on the system where the quark is imbedded.

Since we are mostly interested in the case of light quarks, the
perturbative mass evolution is small and unimportant, and the main
term appears to be $\Delta m^2_{np}$, which we consider now.

Following \cite{10} we consider the quadratic in $(\sigma F)$ term
in (\ref{8}) and expand the exponent to make explicit the
resulting SE term (which is exponentiated after all, yielding
additive contribution to the  Hamiltonian). One has
\be
\lan W_{\sigma\sigma'} (x,y) \ran \cong\left \lan
\left(1+\frac{g^2}{2} \sigma_{\mu\nu} \sigma_{\rho\lambda}
\int^s_0 d\tau \int^s_0 d\tau'F_{\mu\nu} (z(\tau)) F_{\rho\lambda}
(z(\tau'))\right) W_0 +...\right\ran\label{15}\ee where we have
neglected the term proportional to $ds_{\mu\nu} ds_{\lambda\rho}$
since it contributes to the $q\bar q$ potential accounted for in
Hamiltonian and not one-body SE terms; also the mixed  terms
($\sim ds_{\mu\nu} \sigma_{\lambda\rho})$ contribute to the
spin-orbit potentials, also taken into account in the Hamiltonian
\cite{11}. Here $W_0$ is the usual Wegner-Wilson loop without
$(\sigma F)$ operators. The vacuum averaging in (\ref{15}) yields
in the Gaussian approximation (see Appendix of \cite{11})
\be
\lan F_{\mu\nu} F_{\rho\lambda} W_0\ran = \{[\lan F_{\mu\nu}
F_{\rho\lambda}\ran - g^2 \int d s_{\lambda\beta} \lan F_{\mu\nu}
F_{\alpha\beta} \ran \int ds_{\gamma\delta} \lan F_{\rho\lambda }
F_{\gamma\delta}\ran ] \lan W_0\ran \}.\label{16}\ee

Introducing scalar functions $D$ and $D_1$, as in \cite{12} (we
omit for simplicity parallel transporters $\Phi(u,v))$ $$ g^2\lan
F_{\mu\nu} (n) F_{\rho\lambda} (v) \ran = \hat 1 \{
(\delta_{\mu\rho} \delta_{\nu\lambda} - \delta_{\mu
\lambda}\delta_{\nu\rho}) D(u-v)+ $$
\be
+\frac12 [ \partial_\mu (h_\rho \delta_{\mu\lambda} - h_\lambda
\delta_{\nu\rho} ) + \partial_\nu ( h_\lambda\delta_{\mu\rho}
-h_{\rho} \delta_{\mu\lambda})] D_1 (u-v)\}\label{17} \ee with
$h_\mu=u_\mu-v_\mu$, one has $$ \sigma_{\mu\nu}
\sigma_{\rho\lambda} \lan F_{\mu\nu}(z)  F_{\rho\lambda}(z')
W_0\ran = 6 [D (z-z') +D_1(z-z')]- $$
\be
-4\int \sigma_{\alpha\beta}  ds_{\alpha\beta} (u) D (u-z)
\int\sigma_{\gamma\delta} d s_{\gamma\lambda}(v)
D(v-z'),\label{18}\ee where one should have in mind that in
$ds_{\alpha\beta}$ it is always implied $\alpha<\beta$ both in
(\ref{16}) and consequently in (\ref{18}).

Using now the identities
\be
(Dz)_{xy} =(Dz)_{xu} d^4u (Dz)_{uv} d^4v(Dz)_{vy}\label{19} \ee
\be
\int^\infty_0ds \int^s_0 d\tau_1\int^{\tau_1}_0 d \tau_2  f
(s,\tau_1, \tau_2)= \int^\infty_0 ds \int^\infty_0 d \tau_1
\int^\infty_0 d\tau_2 f (s+\tau_1+\tau_2, \tau_1+\tau_2,
\tau_2)\label{20}\ee one can rewrite (\ref{5}), (\ref{7}) with
insertion of (\ref{15}) as $$
 G_{q\bar q} (x,y) = G^{(0)}_{q\bar
q} (x,y)+$$\be+tr [\Gamma (m-\hat D) \Delta_{xu} \sigma_{\mu\nu}
d^4u \Delta_{uv} \sigma_{\rho\lambda} d^4 v \Delta_{vy}
\bar\Gamma(m'-\hat D') \Delta_{yx} \lan F_{\mu\nu} (u)
F_{\rho\lambda} (v) W_0\ran ] \label{20}\ee where we have defined
\be
\Delta_{xu} \equiv\int^\infty_0 ds e^{-K(s)} (Dz)_{xu},~~ K(s)
\equiv m^2s +\frac14\int^s_0 \left( \frac{dz_\mu}{d\tau}\right)^2
d\tau \label{21}\ee An alternative derivation is given in
\cite{10}. Note that $\lan W_0\ran$ depends on trajectories
entering in $\Delta_{zz'}$ in (\ref{20}). One can now take into
account that when $|x-u|$ is small, i.e. $|x-u|\la T_g$ the
influence of $\lan W_0\ran $ on $\Delta_{xu} $ can be neglected,
since $\lan W_0\ran$ is a smooth function of its boundaries,
varying when they are deformed at the scale larger than $T_g$,
while $\Delta_{xu} $ is singular for small $|x-u|$. Indeed in this
limit neglecting the presence of $\lan W_0\ran$ one obtains
\be
\Delta_{xu}^{(0)} = \frac{m}{4\pi^2}
\frac{K_1(m|x-u|)}{|x-u|}.\label{22}\ee

For large $|x-u|$ one can use the fact, that the product of the
spinless quark Green's function  $\Delta_{xy}$ and that of the
spinless antiquark $\Delta_{yx}$ together with $\lan W_0\ran$
yield the asymptotics of the total  meson mass $M_0$ without spin
contributions, and without self-energy corrections
\be
\int\Delta_{xy} \Delta_{yx} \lan W_0\ran \sim \exp
(-M_0|x-y|).\label{23}\ee Therefore we shall use for $\Delta_{xy}$
at large $|x-y|$ the interpolation form
\be
\Delta_{xu} (x)  \cong\frac{\bar m K_1(\bar
m|x|)}{4\pi^2|x|},~~\bar m= m+\tilde M_0,~~ \tilde M_0\approx
\frac{M_0}{2}.\label{24}\ee (We do not need a high accuracy of
(\ref{24}) since it will enter in the small correction term).

Now inserting (\ref{18}) into (\ref{20}) one obtains the following
combination. $$ J( u, v) = 6 [D(u-v)+D_1(u-v)] \Delta_{uv}-$$
\be-4 \sigma_{\alpha\beta} \sigma_{\gamma\delta} \int
ds_{\alpha\beta}{(z)} D(z-u) \int ds_{\gamma\delta} (w)
D(w-v).\label{25}\ee In  the first term on the r.h.s. of
(\ref{25}) $\Delta_{uv}$
 enters with the factors $D(u-v),$ $ D_1(u-v)$ which fall off with
 small correlation length $T_g$. Therefore in \cite{10}
 $\Delta_{uv}$ was taken as $\Delta_{uv}^{(0)}$. Here we tend to
 improve this result by taking into account the asymptotic
 fall-off of $\Delta_{uv}$ as  in (\ref{23}), (\ref{24}). This can
 be done replacing in the free propagator  by the $\Delta_{wv}$ from (\ref{24})
  so that the asymptotics  both at
 small and large distances is reproduced.

 Identifying $m^2_{np}$ from the expansion
 \be
 (m^2+\Delta m^2_{np} - D^2)^{-1}= (m^2-D^2)^{-1}-
 (m^2-D^2)^{-1}\Delta m^2_{np} (m^2-D^2)^{-1}+...\label{26}\ee
 one obtains
 \be
 \Delta m^2_{np} =-\int d^4w\frac{\bar mK_1(\bar m|w|)}{4\pi^2|w|}
 6 (D(|w|) +D_1(|w|))+\sigma^2 \int\Delta_{xu} d^4(x-u),
 .\label{27}\ee
 We take the lattice estimate (for the quenched case \cite{45}),
 $D_1\cong \frac13 D$ and the relation  \cite{12}
$ \sigma=\frac12\int D(z) d^2 z,$ and obtain
\be
\Delta m^2_{np} =-
\frac{4\sigma}{\pi}\varphi(t)+\frac{\sigma^2}{2(m+\tilde
M_0)^2};~~ t=(m+\tilde M_0) T_g. \label{28}\ee

Here $\varphi(t)$ is given in Appendix 2; it is normalized as
$\varphi(0)=1$. The first term on the r.h.s. of (\ref{28})
coincides with the result \cite{10} when $\tilde M_0$ is neglected
(note however that coefficients before $D$ and $D_1$  in \cite{10}
have a misprint, and should be replaced by those in (\ref{25})),
while the last term in (\ref{28}), which is a correction to the
first, is new. To understand the role of this term in creating the
total mass of the meson (still without Coulomb correction and
spin-dependent terms), we take the case of the heavy-light meson,
i.e. when  the quark is moving in the field of an infinitely heavy
antiquark. The Hamiltonian in this case is written as in
\cite{19}-\cite{21} with the SE term treated as in  \cite{10},
namely
\be
H_0(\omega) = \frac{m^2+\Delta m^2_{np}}{2\omega} +
M_0(\omega)\label{27a}\ee where $\tilde M_0(\omega)= \frac{ \omega
}{2}+\varepsilon (\omega), \varepsilon(\omega) =
\frac{\sigma^{2/3} a(n)}{(2\omega)^{1/3}} ; a(0)=2.338$. Taking
into account (\ref{28}) the resulting expression for $
H_0(\omega)$ is \be H_0(\omega)=\left(-\frac{4\sigma
\varphi(t)}{\pi} + \frac{\sigma^2}{2(\tilde M_0+m)^2}\right)
\frac{1}{2\omega}+ \tilde M_0(\omega).\label{28a}\ee Now $\omega$
should be found from the equation \cite{4,7,32}
 \be \frac{\partial
H_0(\omega)}{\partial\omega}|_{\omega=\omega_0}=0\label{29a}\ee

Neglecting in (\ref{29a})  the contribution of the SE term, one
obtains the values of $\omega_0$ and $\tilde M_0, H_0(\omega_0)
\equiv M$ as in \cite{4} which are shown in Table 1  all masses
are given in GeV.

\vspace{1cm}

{\bf Table 1 ~~~Mass eigenvalues (in GeV) according to Eqs.
(\ref{28a}), (\ref{29a}) for different values of $\alpha_s$ }\\

\begin{center}

\begin{tabular}{|l|l|l|l|l|l|l|} \hline
$\alpha_s$& $\omega_0$& $\varepsilon$& $\tilde M_0$& $\varphi(t)$&
$\Delta M_{SE}$& $M=\tilde M_0 + \Delta M_{SE}$\\ \hline 0&0.448&
0.735& 0.96& 0.416& -0.106& 0.854\\ 0.3&0.546& 0.498& 0.771& 0.5&
-0.105& 0.666\\ 0.39& 0.594& 0.407& 0.704& 0.525&-0.101& 0.603\\
\hline
\end{tabular}
 \end{center}
For $\alpha_s>0$ the  values $\varepsilon (\omega),$  $\tilde M_0
=\frac{\omega}{2}+ \varepsilon (\omega)$ have been calculated in
\cite{4} taking  the color Coulomb term $-\frac43
\frac{\alpha_s}{r}$ into  account, while $\Delta M_{SE}  =  -
\frac{2\sigma}{\pi\omega_0} \varphi (t)$  and $ t=T_g \tilde M_0$,
$T_g=1$ GeV$^{-1}$. One can see that $\Delta M_{SE} $ is rather
stable and gives a correction around 15\% to the total mass. The
correction of the second term on the r.h.s. of  (29) to the total
$\Delta m^2$ is of the order of 7\% for $m=0$ , so that the
earlier calculations made without this term in \cite{14}-\cite{18}
and \cite{22}-\cite{24} would be modified by few percent.

As the next comparison one can take the solution of Dirac equation
for the heavy-light meson with confining and color Coulomb term
present. In this  case the SE term is absent in the first order
Dirac Hamiltonian  (in contrast to the second-order SA
Hamiltonian, obtained from FFSR). The results of calculations,
performed  in \cite{47,48} are shown in Table 2. In this case to
compare  with the SA Hamiltonian (\ref{28a}) one should take $T_g$
in $\varphi (t)$ equal to zero, since the  linear confining
potential
 is obtained in this limit (while for $T_g\neq 0$ there
appear corrections to the linear potential calculated in \cite{48}).
 Hence in (\ref{28a})  one puts $\varphi(t=0)=1$ and neglects as
 before  the second correction term in brackets on the r.h.s. As a
 result one obtains for $\sigma$=0.16 GeV$^2$ (all masses are
 given in GeV).

\vspace{1cm}

{\bf Table 2 Masses and self-energy corrections according to Eq.
(\ref{28a}) in comparison with eigenvalues of Dirac equation. }\\

\begin{center}

\begin{tabular}{|l|l|l|l|} \hline
$\alpha_s$& 0&0.3&0.39\\ \hline
 $\Delta M_{SE}$& -0.227&-0.186&-0.171\\
 \hline
 $M=\tilde M_0 + \Delta
M_{SE}$& 0.733&0.585&0.533\\ \hline $M_D$& 0.65&0.465&0.401\\
\hline
\end{tabular}
 \end{center}

In the last line the Dirac eigenvalues from \cite{47,48} are given
to be compared with the eigenvalues of (\ref{28a}) in the next
line. One can see that Dirac eigenvalues are about 100 MeV lower.
This difference can be attributed to the fact that in Dirac
equation both positive and negative eigenvalues (the latter
corresponding to the backward-in-time motion of quark) are taken
into account, while  in (\ref{28a}) only positive values of
$\omega_0$ are considered. In the next section we shall discuss
how the negative modes (negative solutions for $\omega_0$) can be
included in the SA Hamiltonian.

\section{The matrix Hamiltonian for the heavy-light $q\bar q$ system}

We start with the Hamiltonian for the free Dirac particle $\hat H=
m\beta + \veal \vep$, which can be diagonalized using
Foldy-Wouthuyzen (FW) procedure
\be
\hat H= U^+\hat H_d U,~~ \hat H_d =\left( \begin{array}{ll}
\sqrt{\vep^2+m^2}&0\\
0&-\sqrt{\vep^2+m^2}\end{array}\right).\label{4.1}\ee

As it is explained in detail in Appendix 1 the free Green's
function can be written in terms of $\hat H_d$ as
\be
S(t)=i\beta U\left( \begin{array}{ll} \theta(t)&0\\
0&-\theta(-t)\end{array}\right) e^{-i\hat
H_{d}t}U^+.\label{4.2}\ee

In the Appendix 1 also the case of the Weyl representation is
discussed for the Dirac particle in the external fields, which
gives a representation similar to (\ref{4.2}), i.e. having the
diagonal Hamiltonian of the form of $\hat  H_d$ in  the exponent
for the time-dependent Green's function.

We now turn to the FFS form of the quark Green's function in the
external gluonic field, written with the help of the einbein
function $\omega$ \cite{4,7} (this function was previously denoted
as $\mu$ in most papers) \be S_q(x,y) = \int D\omega
(D^3z)_{\vex\vey} e^{-\int^T_0 \left (
\frac{m^2}{2\omega}+\frac{\omega}{2} + \frac{\omega \dot
z^2_i}{2}\right)  dt + ig \int^x_y A_\mu dz_\mu + g \int^T_0
\sigma F\frac{ dt}{2\omega}}.\label{4.3}\ee After vacuum averaging
this function can be associated with the Green function of the
heavy-light meson.

Here $D\omega$ is the  path integration over functions $\omega
(t)$, which in our formalism  \cite{4,7} is calculated by the
stationary point (steepest descent) method, after going over  to
the Hamiltonian form instead of the Lagrangian  path integral form
of (\ref{4.3}), namely,
\be
S_q(x,y) = \left \lan  x\left | \int D\omega e^{-i\int_0^{T_M}
\left(H_0(\omega) - g\frac{\sigma F}{2\omega}\right)
dt_M}\right|y\right \ran,\label{4.4}\ee where we have changed from
the Euclidean time $t$ to the Minkowskian time  $t_M=-it$,  and
\be
H_0(\omega) = \frac{m^2}{2\omega} +\frac{\omega}{2} +
\frac{\vep^2}{2\omega} + \sigma r.\label{4.5} \ee Solving the
Schroedinger-type equation
\be
\left (\frac{\vep^2}{2\omega} +\sigma r\right) \varphi_n =
\varepsilon_n{(\omega)} \varphi_n,\label{4.6}\ee one obtains
\be
\varepsilon_n(\omega) = \frac{\sigma^{2/3}}{(2\omega)^{1/3}}
a_n\label{4.7}\ee
 where $a_n, n=0,1,2,...$ is the set of zero of Eiry functions,
 $a_0\cong  2.338$.
As a result for $L=0$ one has the eigenvalues $E_n^{(0)}(\omega)$
of $H_0(\omega)$, equal to \be E_n^{(0)}(\omega) =
\frac{m^2}{2\omega} + \frac{\omega}{2} +
\frac{\sigma^{2/3}}{(2\omega)^{1/3}} a_n.\label{4.8} \ee In our
previous calculations of the spectrum the stationary point in the
integration over $D_\omega$ was taken at the positive solution of
the  equation
\be
\frac{\partial E_n^{(0)} (\omega)}{\partial \omega} |_{\omega=
\omega_n^{(0)}}=0, ~~ \omega_n^{(0)}>0.\label{4.9}\ee

However there is a negative solution, at
$\omega_n=-\omega_n^{(0)}$, which was neglected  in all previous
calculations.

The Hamiltonian (\ref{4.5}) refers actually to the gauge-invariant
system of a heavy-light $q\bar Q$ system, where the infinitely
heavy quark $\bar Q$ propagates along the time axis and is
situated at the spacial origin.

In line with the Hamiltonian (\ref{4.1}), we can write  the total
Hamiltonian of the  $q\bar Q$ system with the lower Dirac
components as \be \hat H_{q\bar Q}=\left(\begin{array}{ll}
h_0&h_{+-}\\h_{-+} &-h_0\end{array}\right), h_0\varphi
=E^{(0)}_n\varphi \label{4.10}\ee In  $h_0$ the spin-dependent
term $\frac{g(\sigma F)}{2\omega}$ in the exponent of (\ref{4.4})
 does not contribute to the diagonal part of (\ref{4.10}) for
$s$-wave states of heavy-light mesons, except for the diagonal SE
term considered in the previous section (this term can be added
replacing $m^2$ in (\ref{4.5}), (\ref{4.8}) by $m^2+\Delta
m^2_{np}$).

Now we turn to the calculation of the terms $h_{+-}= h_{-+}^*$. To
this end one can use Eqs.(\ref{5}), (\ref{6}), (\ref{8}) having in
mind that for the heavy-light meson the Green's function of the
heavy quark reduces to the parallel transporter $\Phi(x,y)$ along
the straight line and the factors $ds'(Dz')_{xy} e^{-K'}$ and
absent. Having in mind (\ref{4}) one must calculate the factor
\be
\lan W_{\sigma\sigma'}^{(2)} (x,y) \ran =\exp \left [
-\frac{g^2}{2}\int d\pi(1) d\pi(2) \lan F(1) F(2)\ran
\right].\label{43} \ee In the product $d\pi(1) d\pi(2)$ the term
$ds(1) ds(2)$ contributes to the linear interaction and is present
in $H_0(\omega), $ the term $d\tau(1) d\tau(2)$ was calculated in
the previous section and in \cite{10} and was taken into account
in $\Delta m^2_{np}$. The mixed terms can be written as
\be
\exp\left (-g^2 \int ds_{\mu\nu} (u) d\tau \left \lan F_{\mu\nu}
(u) \left ( \begin{array}{ll} \vesig \veB & \vesig \veE\\ \vesig
\veE& \vesig \veB \end{array} \right)_{z(\tau)} \right\ran
\right).\label{44}\ee The diagonal terms in (\ref{44}) contribute
to the spin-orbit interaction, computed  in \cite{11}, and vanish
for the $s$-states, while the nondiagonal terms contribute to
$h_{+-}, h_{-+}$ and calculated below. Writing \be ds_{\mu\nu}
F_{\mu\nu} (u) = ds_{i4} E_{i} (u) + ds_{ik} F_{ik} = n_{i} d^2 u
E_i(u) + d\ves \veB,~~ \ven = \frac{\veu}{|\veu|},\label{45}\ee
 one should average this term with nondiagonal component $\vesig
                          \veE$
 and neglect the last term, since the correlator (\ref{45})
 $\lan B_i E_k\ran$ is proportional to $\frac{\partial
 D_1}{\partial u_l}$ and small, as a result one obtains the
 integral
 \be
 \int d^2 u \lan (\ven \veE (u) ) (\vesig \veE (z(\tau))\ran =
 \vesig \ven \int d^2 u  D(u-z) = \vesig \ven \sigma \label{46}\ee
 where we have taken into account the definition  $\sigma =\frac12
 \int d^2u D(u)$, and the fact that $z(\tau)$ lies on the quark
 trajectory, which is the boundary of the integration surface.

Finally one must replace $d\tau$ in (\ref{44}) by $dt$, having in
mind that upper matrix elements refer to the positive time
evolution, while lower ones (corresponding to the negative energy
eigenvalues) refer to the negative (backward) in time  evolution,
viz.
\be
\int d\tau \left ( \begin{array}{ll} a&c\\ d& b\end{array} \right)
= \left ( \begin{array}{rr} \int\frac{dt}{2\omega}
a&\int\frac{dt}{2\omega}c \\ -\int\frac{dt}{2\omega} d&-\int
\frac{dt}{2\omega} b\end{array} \right).\label{47} \ee As a
consequence from (\ref{44}) and (\ref{46}) one obtains
\be
h_{+-} = \frac{i\sigma}{2\omega} (\vesig \ven),~~ h_{-+} =
-\frac{i\sigma}{2\omega} (\vesig \ven).\label{48}\ee

The energy eigenvalues of the matrix Hamiltonian ({42}) are
obtained in the usual way from the equation
\be
\det \left( \begin{array}{ll} h_0-E, & h_{+-}\\ h_{-+},&
h_0-E\end{array}  \right)  = 0,~~ h_0\equiv E_n^{(0)} (\omega)
\label{49}\ee which yields
\be
E=\pm \sqrt{ h_0^2+\left(
\frac{\sigma}{2\omega}\right)^2}\label{50}\ee  where  $\omega$
should be found from the condition $\frac{\partial
E}{\partial\omega} |_{\omega=\omega_0}=0$, which replaces the old
condition (\ref{4.9}), and can be rewritten as
\be
2h_0h'_0-\frac{\sigma^2}{2\omega_0^3} =0.\label{51}\ee

Writing $h_0 = \frac{m^2-\Delta}{2\omega} + M_0 (\omega)$ , with
$\Delta = \frac{4\sigma}{\pi} \varphi(t),$ one can rewrite
(\ref{51}) for the case $m=0 ,~~ T_g=0,~~ \varphi\equiv 1,$
\be
\omega_0 =\bar \omega_0 \left\{ 1- 2\left(
\frac{c_1}{\omega_0^2}-\Delta \right) \left(\frac{\omega_0}{\bar
\omega_0}\right)^{4/3} \right\}^{3/4}\label{52}\ee where $c_1
=\frac{2\sigma}{\pi}, ~~ \bar \omega_0 =\sqrt{2\sigma}
\left(\frac{a}{3}\right)^{3/4},\Delta =\frac{\sigma^2}{4\omega^3_0
h_0 (\omega_0)}.$

Solving (\ref{52}) one obtains for $\sigma=0.18$ GeV$^2$
 \be
 h_0(\omega_0)\approx0.56 {\rm~ GeV} ,~~ \omega_0 \approx 0.21 {\rm~ GeV}\label{53}\ee and the energy eigenvalue
(\ref{50}) equal  to
\be
E_0=E(\omega_0) =\pm 0.70 {\rm~GeV}\label{54}\ee which should be
compared with the Dirac eigenvalue from Table 2, $
E_D=1.619\sqrt{\sigma}=0.686 {\rm ~ GeV}.$ Thus taking into
account matrix structure of the  Hamiltonian diminishes eigenvalue
by $\sim 0.1$  GeV and yields values  in good agreement with
independent calculation of Dirac equation.

It is of interest to compare this Hamiltonian with the Hamiltonian
of the Dirac equation for a light quark in the static source of
linear potential, considered in \cite{47,48}.
\be
\hat H_{Dirac} = \veal\vep + \beta (m+\sigma r).\label{4.11}\ee
For the solution $\psi_n(\ver)$ represented in the form \cite{47}
\be
\psi_n(\bar r) =\frac{1}{r} \left( \begin{array}{l}
G_n(r)\Omega_{jlM}\\ iF_n(r) \Omega_{jl'M}\end{array}
\right)\label{4.12} \ee the equation $H_{Dirac}
\psi_n=\varepsilon_n\psi_n$ assumes the form
\be
\left\{ \begin{array}{l} \frac{dG_n}{dr}+\frac{\kappa}{r} G_n-
(\varepsilon_n+m+\sigma r) F_n=0\\
\frac{dF_n}{dr}+\frac{\kappa}{r} F_n- (\varepsilon_n-m-\sigma r)
G_n=0\end{array} \right. \label{4.13} \ee where
$\kappa(j,l)=(j+\frac12) sgn (l-j)$.

Equations (\ref{4.13}) are invariant under the  substitution
($\varepsilon_n, G_n, F_n, \kappa) \leftrightarrow
(-\varepsilon_n, F_n, G_n, -\kappa)$. This means that for every
solution with $\varepsilon_n>0$, and $\kappa$ having the form
(\ref{4.12}) there exists another solution of the form
\be
\psi_{-\varepsilon_n} (r) =\frac{1}{r} \left( \begin{array} {l}
F_n(r) \Omega_{jl'M}\\ iG_n(r) \Omega_{jlM}\end{array} \right)
\label{4.14}\ee which has eigenvalue $-\varepsilon_n, -\kappa$.

Following the idea  of the FW transformation leading to
(\ref{4.1}) one can also assume that the Hamiltonian (\ref{4.11})
can be diagonalized to the form $$ \hat H_{Dirac} \to U^+\left(
\begin{array}{ll}\hat h(\kappa)&\\&-\hat h(-\kappa)\end{array} \right) U,
$$ where $\hat h(\kappa) \varphi^{\kappa}_n=\varepsilon_n
\varphi^{\kappa}_n $ and $\hat h(-\kappa)
\varphi^{-\kappa}_n=\varepsilon_n \varphi^{-\kappa}_n $.

This brings us to the eigenvalue matrix (\ref{4.10}). One special
feature of this representation is that the states
$\psi_{\varepsilon_n}$ and $\psi_{-\varepsilon_n}$ have different
parities.

\section{Negative energy states for the $q\bar q$ mesons}

We are now considering the $q\bar q$ meson states made of light
quarks. The Hamiltonian for positive energy states was used
repeatedly (see \cite{32} for details) and has the form
\be
H_0 (\omega_1, \omega_2) =\frac{m^2_1- \Delta_1}{2\omega_1} +
\frac{m^2_2-\Delta_2}{2\omega_2} + \frac{\omega_1+\omega_2}{2} +
\frac{\vep^2}{2\tilde \omega} + \sigma  r\label{5.1}\ee where
$\tilde \omega=\frac{\omega_1 \omega_2}{\omega_1+\omega_2},~~
\Delta_i = \frac{4\sigma}{\pi} \varphi(t_i),~~ t_i = (m_i +\tilde
M_0) T_g.$ Turning now to the spin-dependent term ($\sigma F)$ in
(35) and (\ref{8}), one remarks that in addition to the one-quark
corrections considered in the previous section one has also the
spin-spin term, previously treated in \cite{11} and yielding the
hyperfine interaction, namely the term $V_4$. However in the
derivation only the diagonal components of the matrix $\lan
\sigma_{\mu\nu}^{(1)} F_{\mu\nu} \sigma_{\alpha\beta}^{(2)}
F_{\alpha\beta}\ran$ have been taken into account and now we shall
look carefully into the nondiagonal terms.

From (\ref{8}) one has the following contribution
 (Note that for antiquark the spin operator $\sigma^{(2)}_{\mu\nu}$
  enters with the spin opposite to $\sigma^{(1)}_{\mu\nu}$)
 \be
\exp \left\{ -\frac12 \int^{s_1}_0 d\tau_1 \int^{s_2}_0 d\tau_2
g^2 \left \lan\left ( \begin{array}{ll}\vesig^{(1)} \veB&
\vesig^{(1)} \veE\\\vesig^{(1)} \veE& \vesig^{(1)}
\veB\end{array}\right)_{z(\tau_1)} \left ( \begin{array}{ll}
\vesig^{(2)} \veB& \vesig^{(2)} \veE\\\vesig^{(2)} \veE&
\vesig^{(2)} \veB\end{array}\right)_{z(\tau_2 )}\right \ran
\right\}\label{5.2}\ee and one should replace as usual $d\tau_i
=\pm\frac{dt_i}{2\omega_i}$.

As a result on obtains the following spin-spin terms in the
Hamiltonian: i) from the product of diagonal components $\lan
\vesig^{(1)} \veB\vesig^{(2)} \veB\ran$ one has the usual
hyperfine interaction \cite{11,32}\be \hat V_{hf}^{(diag)}(r)
=\frac{\vesig^{(1)}\vesig^{(2)}}{12\omega_1\omega_2}
\int^\infty_{-\infty} d\nu \left[ 3D(r,\nu) + 3D_1(r,\nu) +
2\ver^2 \frac{\partial D_1}{\partial r^2}\right]\label{61}\ee
where $D(r,\nu)=D(\sqrt{r^2+\nu^2})$, and
$\ver=\vez_1(t)-\vez_2(t)$ is the quark-antiquark distance, and we
have used (\ref{17}) to calculate $\lan B_i(u)B_k(v)\ran$. Eq.
(\ref{61}) contains both perturbative and nonperturbative
contributions, the latter have been calculated in \cite{11} and
found to be much smaller than the perturbative ones, which can be
easily calculated using the lowest order form of $D_1$ \cite{11}
\be
D_1^{pert} (x) =\frac{16}{3\pi} \frac{\alpha_s}{x^4}
+O(\alpha^2_s)\label{62}\ee which gives the standard result
\be
\hat V_{nf}^{(diag)} = \frac{8\pi \alpha_s
\vesig^{(1)}\vesig^{(2)}\delta^{(3)}(\ver)}{9\omega_1\omega_2}.\label{63}\ee
The matrix element of (\ref{63}) can be written as
\be
\lan\hat V_{nf}^{(diag)}\ran = \frac{2 \alpha_s \lan
\vesig^{(1)}\vesig^{(2)}\ran}{9\omega_1\omega_2}R_n^2(0)=
\frac{4\alpha_s\tilde
\sigma}{9(\omega_1+\omega_2)}\left(\begin{array}{ll}-3,&S=0\\+1,&S=1\end{array}\right)\label{64}\ee
where $\tilde \sigma \equiv \sigma+\frac43 \alpha_s \lan
r^{-2}\ran$, for more details see Appendix  3 of \cite{37}.

 ii) We now turn to the product of nondiagonal terms in (\ref{5.2}),
which can be written similarly to (\ref{61}) as
\be
\hat V^{(nond)}_{hf} (r) =\frac{\vesig^{(1)}
\vesig^{(2)}}{12\omega_1\omega_2}\int^\infty_{-\infty} d\nu \left[
3D + 3D_1 + (3\nu^2+r^2) \frac{\partial D_1}{\partial
r^2}\right].\label{65}\ee Keeping again only perturbative
contribution, one easily obtains \be \hat V_{hf}^{(nond)} =-\hat
V_{hf}^{(diag)}.\label{66}\ee Consider now the total Hamiltonian
\be
\hat H=(H_0(\omega_1,\omega_2) + \hat V_{hf}^{(diag)}) \hat
1_1\hat1_2 + \hat V_{hf}^{(nond)} (\gamma_5)_1
(\gamma_5)_2.\label{67}\ee The energy eigenvalues can be found in
the same way, as it was done in the previous section
\be
E^{(\omega_1,\omega_2)} = \pm \sqrt{\left(H_0(\omega_1, \omega_2)+
\hat V^{(diag)}_{hf} \right)^2+\left(\hat
V^{(nond)}_{hf}\right)^2}.\label{68}\ee

To illustrate the general result (\ref{68}) we take the case of
massless quarks, and write the energy as
\be
E(\omega) =\pm \sqrt{\left( h_0 (\omega) +\frac{c_\sigma}{\omega}
\right)^2 + \frac{c_\sigma^2}{\omega^2}}\label{69}\ee where
$h_0(\omega)$ is the eigenvalue of $H_0(\omega_1, \omega_2),
\omega_1= \omega_2$
\be
h_0 (\omega) =-\frac{\delta}{\omega} + \omega +
\frac{c}{\omega^{1/3}}; c=\sigma^{2/3} a(n), a(0) =
2.338.\label{70}\ee

Also we have defined
\be
c_\sigma= \frac{2\alpha_s\tilde \sigma}{9} \left(\begin{array}{ll}
-3,&S=0\\1,&S=1\end{array} \right).\label{71}\ee

The crucial point is now that $\omega$ is to be  found as before
from the minimum of $E(\omega)$ (we assume that minimization of
the eigenvalue $E(\omega)$ instead of the operator Hamiltonian
$H(\omega_1,\omega_2)$ brings about a small correction as it was
in the case of the one-channel Hamiltonian, see \cite{7} and
numerical analysis in the second ref. of  \cite{7}).  This section
served as an illustration of positive-negative state mixing due to
hyperfine interaction in $\bar q q$ mesons. For the lack of space
the detailed analysis of the corresponding change in the spectrum
will be published elsewhere.

\section{ Conclusions}

In this paper the systematic discussion is started of the role of
negative energy components (NEC) for quark bound states. The NEC
are automatically taken into account in the one-body Dirac or
Bethe-Salpeter formalism. In the  latter case however the
Bethe-Salpeter wave-function contains for the $q\bar q$ meson at
least eight independent components and their relative role can be
studied only numerically. A didicated analysis of NEC was done in
the quasipotential approximation of the Bethe-Salpeter equations
\cite{49} and effects of NEC was found to be significant for the
spectrum  of mesons.  Recently another approach was introduced in
\cite{50,48} and developed further in \cite{51,52},  called the
Method of Dirac Orbitals (MDO), where the quark bound state is
expanded in a series of products of individual one-body Dirac
states. In this case also the NEC effects are taken into account,
but other approximations are usually done (c.m. motion, higher
components) which require a cross-check of all results and
comparison to other formalisms.

The Hamiltonian formalism and in particular the SA Hamiltonian is
physically transparent and  mathematically simple,  it reduces to
the popular Relativistic Quark Model (RQM) Hamiltonian in its
simplest form (when string motion is neglected) and therefore it
is necessary to understand the role of NEC in the Hamiltonian
form.

This  is done in the present paper using the simplest bound system
-- the heavy-light meson, where the heavy quark plays the role of
external field and therefore results can be compared  to those of
Dirac equation. The comparison proceeds in two steps. Firstly one
takes in the  Hamiltonian the one-body self-energy terms which do
not mix  positive energy components and  NEC. Here a new
correction term was obtained in section 3, Eq. (41) in addition to
the old one \cite{10} which  gives around 7\% of the total.
Secondly, the NEC mixing appears due to the nondiagonal
Hamiltonian matrix elements. The stationary point condition for
the einbein variable $\omega$ should now be applied to the
eigenvalues of the total matrix Hamiltonian $E(\omega)$ \be \hat
H= \left(
\begin{array}{ll} h_0& h_{+-}\\h_{-+} & -h_0\end{array}\right), ~~
det (\hat H - E(\omega)) =0, ~~ \frac{\partial E(\omega)}{\partial
\omega} =0.\ee

The resulting stationary values $\omega_0$ and $E(\omega_0)$ are
in good agreement with the eigenvalue of the Dirac operator $E_D$.

\be E(\omega_0) \cong E_D.\ee

This procedure justifies the stationary  analysis with respect to
$\omega$, since for $h_0(\omega)$ the stationary point does not
exist for light quarks if the self-energy term $-\frac{\Delta
m^2}{2\omega}$ is included in  $h_0(\omega)$, while for
$E(\omega)$ the stationary point $\omega\equiv \omega_0$ always
exists. In chapter 5 the first step is done for arbitrary $q\bar
q$ systems, and the matrix Hamiltonian is written down explicitly.
The main lesson here is that NEC are mixed up by the  hyperfine
interaction and the doubly nondiagonal (for both quark and
antiquark) terms can be calculated explicitly. The further
analysis and numerous applications are relegated for the lack of
space  to future publications. However already at this stage it is
clear that NEC are very important for the structure of the
wave-functions and eigenvalues of mesons. This is probably even
more so for baryons, where NEC are responsible for the correct
relativistic structure  of baryon wave-functions, which is clearly
seen in the values of the $g_A/g_V$ ratio \cite{53}.

The author is grateful to A.B.Kaidalov, I.M.Narodetsky, J.A.Tjon,
M.A.Trusov and J.Weda and members of ITEP seminar theory group for
useful  discussions.

The work  is supported
  by the Federal Program of the Russian Ministry of industry, Science and Technology No.40.052.1.1.1112.\\

\newpage

 \setcounter{equation}{0}
\renewcommand{\theequation}{A.\arabic{equation}}

\begin{center}
{\bf Appendix 1}\\

{\large }
\end{center}

In this appendix we shall derive several representations for the
quark Green's function  in the external nonabelian field. We start
with the standard  Fock-Feynman-Schwinger Representation (FFSR) as
a warm up. To this end one writes first the proper-time
representation in the Euclidean space-time
\be
S=(m+\hat D)^{-1}= (m-\hat D)(m^2-\hat D^2)^{-1}= (m-\hat D)
\int^\infty_0 ds e^{-s(m^2-\hat D^2)}.\label{A.1}\ee

Now one transforms (\ref{A.1}) to the path-integral as follows
\be
\lan x|\int^\infty_0 ds e^{s D_\mu^2}|y\ran = \lan
x|e^{\varepsilon D^2_\mu(N)}|x_{n-1}\ran \lan x_{n-1}|
e^{\varepsilon(D^2_\mu(N-1))}|x_{n-2}\ran...\lan
x_1|e^{\varepsilon D^2_\mu(1)}|y\ran.\label{A.2}\ee

In (\ref{A.2}) the integration over all $d^4 x_1... d^4 x_{N-1}$
is implied and the relation $s=\varepsilon N$ is used. Consider
now one piece of the path in (\ref{A.2}) and write $$ I_{n,n-1}
\equiv \lan x_n| e^{\varepsilon (\partial_\mu-ig
A_\mu)^2}|x_{n-1}\ran = \lan x_n|p\ran \frac{d^4p}{(2\pi)^4}
e^{\varepsilon (\partial_\mu-ig
A_\mu(\frac{x_n+x_{n-1}}{2}))^2}\lan p|x_{n-1}\ran = $$
\be
=\frac{d^4p}{(2\pi^4} e^{ip(x_n-x_{n-1})-\varepsilon (p_\mu-g
A_\mu(\frac{x_n+x_{n-1}}{2}))^2}.\label{A.3}\ee

Integration over $d^4p$ in (\ref {A.3}) gives
\be
I_{n,n-1}=\frac{1}{(4\pi\varepsilon)^2}e^{-\frac{(\Delta
x)^2}{4\varepsilon}+ig\Delta x_\mu A_\mu}, ~~ \Delta x = x_n -
x_{n-1}.\label{A.4}\ee Insertion of (\ref{A.4}) in (\ref{A.2})
finally yields
\be
S=(m-\hat D) \int^\infty_0 ds e^{-sm^2} (Dz)_{ xy} e^{-\frac{1}{4}
\int^s_0\dot z^2_\mu d\tau +ig \int^x_y A_\mu dz_\mu + g\int^s_0
\sigma_{\mu\nu} F_{\mu\nu} d\tau}\label{A.5}\ee where we have used
the relation $\hat D^2=D^2_\mu +g\sigma_{\mu\nu} F_{\mu\nu}$.

Note that in FFSR (\ref{A.5}) the exponent contains
$\gamma$-martices only in the spin term
$\sigma_{\mu\nu}F_{\mu\nu}$, and moreover it commutes with
$\gamma_5$. Therefore in the chiral limit $(m\to 0)$ $S$ is odd in
$\gamma_\mu$ irrespectively of any vacuum
 averaging of terms containing $A_\mu$ and $F_{\mu\nu}$. Hence in
 this form one cannot describe the effect of the chiral symmetry
 breaking and one should look for other representations which will be the topic of other publications.

 We start with the case of the free quark and write the Green's
 function in the energy and the
 time-dependent representations $S(E)$  and  $S(t)$ (in the Minkowskian space-time)
 \be
 S(E)=\frac{1}{m+i\hat p}= \frac{\beta}{\hat
 H-E}=\frac{\beta}{m\beta+ \veal\vep-E};~~ S(t)
 =\int^\infty_{-\infty} S(E)
 e^{-iEt}\frac{dE}{2\pi}.\label{A.6}\ee
 Consider now the Foldy-Wouthuyzen (FW) transformation of  the
 free Hamiltonian\be
 U^+\hat HU= U^+\left( \begin{array}{ll}m&\vesig\vep\\ \vesig\vep&
 -m\end{array}\right) U\equiv \hat
 H_d=\left(\begin{array}{ll}\sqrt{\vep^2+m^2}& 0\\
 0&-\sqrt{\vep^2+ m^2}\end{array}\right)\label{A.7}\ee
 where
 \be
 U=\left( \begin{array}{ll}\cos\hat \theta&-\sin \hat \theta\\
 \sin\hat \theta&\cos \hat \theta\end{array}\right),~~
U^+=\left( \begin{array}{ll}\cos\hat \theta&\sin \hat \theta\\ -
\sin\hat \theta&\cos \hat \theta\end{array}\right),\label{A.8}\ee
and \be \sin 2\hat \theta= \frac{\vesig\vep}{\sqrt{\vep^2+m^2}},
\cos 2\hat \theta=\frac{m}{\sqrt{\vep^2+m^2}},~~ \hat \theta
=\frac12 \arctan(\frac{\vesig\vep}{m}).\label{A.9}\ee Consequently
one has \be S(E)=\beta U\frac{1}{\hat H_d-E} U^+,~~
S(t)=\frac{\beta }{2\pi}U \int^\infty_{-\infty}\frac{dE(\hat
H_d+E)}{\vep^2 +m^2-E^2} U^+e^{-iEt}\label{A.10}\ee

Integrating in (\ref{A.10}) one gets finally
\be
S(t) = i\beta U\left(
\begin{array}{ll}\theta(t)&\\&-\theta(-t)\end{array}\right) U^+
e^{-i\sqrt{\vep^2+m^2}|t|}.\label{A.11}\ee Another  form can be
given to $S(E)$ using  the proper-time  representation
\be
S(E) = \frac{1}{m+i\hat p} =i\beta \int^\infty_0 e^{-i(m\beta+
\veal\vep +ip_4)s} ds =i\beta U\int^\infty_0 e^{-i(\hat
H_d+ip_4)s} dsU^+.\label{A.12}\ee

The form (\ref{A.12}) is interesting since it contains the matrix
Hamiltonian in the exponent and we shall use it now to take into
account external field $A_\mu$. The form (\ref{A.7}) is especially
convenient in the nonrelativistic case when $|\vep|\ll m$, and
then also $\theta\ll 1$, and the FW transformation is nearly
diagonal. In the opposite case, $|\vep|\gg m$, one needs to start
from another representation of $\gamma$-matrices, namely the Weyl
representation:
\be
(E- \vep\vesig_W-m\gamma_0^{(W)})\psi =0;~~
\vesig_W=\left(\begin{array}{l}
\vesig,0\\0,-\vesig\end{array}\right), \gamma_0^{(W)}=
\left(\begin{array}{l}0~~1\\1~~0
\end{array}\right),\label{A.13}\ee so that the Green's function in
the Weyl representation is \be S_W(E)=(\vesig_W\vep+m\gamma_0
-E)^{-1}.\label{A.14}\ee Doing the FW transformation one has
similarly to (\ref{A.7}) \be U^+_W\hat H_W U_W= \hat
H^{(W)}_d=\left( \begin{array}{ll}
\sqrt{\vep^2+m^2}\frac{\vesig\vep}{|\vep|},&0\\ 0,&
-\sqrt{\vep^2+m^2}\frac{\vesig\vep}{|\vep|}\end{array}\right)\label{A.15}\ee
where \be U_W=\left(\begin{array}{l}\cos\theta_W, -\sin\theta_W\\
\sin\theta_W,\cos\theta_W \end{array}\right) \equiv
e^{-it_2\theta_W}, ~~\theta_W=\frac12 \arctan
\frac{m\vesig\vep}{\vep^2}\label{A.16}\ee and $t_2\equiv \sigma_2$
is the Pauli matrix in the helicity indices.

\newcommand{\vePm}{\mbox{\boldmath${\rm\mathcal{P} }$}}

We now use the proper-time representation for $S_W$
\be
S_W=\lan x| \int^\infty_0 ds e^{is (\vesig_W\vePm+ m\gamma_0+i
{\cal {P}}_4)}|y\ran\label{A.17}\ee where $ \mathcal{P}_\mu
=\frac{1}{i} \partial_\mu-gA_\mu$, and split the interval $(x,y)$
into $N$ steps as in (\ref{A.2}),$N\varepsilon=s$. One has $$
I^{(W)}_{n,n-1}\equiv \lan x_n|e^{i\varepsilon
(\vesig_W\vePm+m\gamma_0+i \mathcal{P}_4)}|x_{n-1}\ran=$$
\be
=\int\frac{d^4p}{(2\pi)^4} e^{ip(x_n-x_{n-1})} U_W(\theta_W{(n)})
e^{i\varepsilon(\hat H_D^{(W)}(n) +i
\mathcal{P}_4(n))}U^+_W(\theta_W(n)).\label{A.18}\ee

As in (\ref{A.3}) one can write $ \mathcal{P}_\mu =p_\mu-gA_\mu$,
and integrate over $d^4p$, representing the square root terms in
$\hat H_d^{(W)}$ through the einbein function $\mu(x_4)$,
\be
e^{i\varepsilon\sqrt{\vep+m^2}}\sim \int d\mu_n e^{i\left (
\frac{\vep^2+m^2}{2\mu_n} + \frac{\mu_n}{2}\right)
\varepsilon}\label{A.19}\ee one   can also use identity;
\be
e^{ia\vesig \ven} =
e^{ia}\frac{(1+\vesig\ven)}{2}+e^{-ia}\frac{(1-\vesig\ven)}{2}.\label{A.20}\ee
Now the integration over $d^4p$ yields
\be
I^{(w)}_{n,n-1} = \int d\mu_n \left(
\frac{\mu_n}{2\pi\varepsilon}\right)^2 U_W(\theta_W)
e^{igA_\mu\Delta x_\mu} G(n) U^+_W(\theta_W)\label{A.21}\ee where
we have defined the diagonal matrix $G(n)$ with elements $$
G_{++}(n)=e^{-\Delta x_4[\frac{\mu}{2}(\dot{
\vex}^2+1)+\frac{m^2}{2\mu}]}\theta(\Delta x_4) \frac{1+\vesig
\ven}{2} +e^{\Delta x_4[\frac{\mu}{2}(\dot
{\vex}^2+1)+\frac{m^2}{2\mu}]}\theta(-\Delta x_4) \frac{1-\sigma
\ven}{2}$$
\be
G_{--}(n)=e^{-\Delta x_4[\frac{\mu}{2}(\dot
x^2+1)+\frac{m^2}{2\mu}]}\theta(\Delta x_4) \frac{1-\vesig
\ven}{2} +e^{\Delta x_4[\frac{\mu}{2}(\dot
{\vex}^2+1)+\frac{m^2}{2\mu}]} \frac{1+\vesig
\ven}{2}\theta(-\Delta x_4)\label{A.22}\ee and
$\dot{\vex}=\frac{\Delta\vex}{\Delta x_4}, ~\Delta
x_\mu=(x_n-x_{n-1})_\mu$, while $\vep$ residing in $\theta_W$ is
$\vep=\mu\dot {\vex}$.

\newpage

 \setcounter{equation}{0}
\renewcommand{\theequation}{A2.\arabic{equation}}

\begin{center}
{\bf Appendix 2}\\

 \vspace{0.5cm}

\end{center}

The function $\varphi(t), t\equiv m/\delta$, defined in Eq.(29)
can be written as (note the difference in definition here and in
\cite{9})
\be
\varphi(t)= t\int^\infty_0 z^2 dzK_1(tz) e^{-z}\label{a.1} \ee
where $K_1$ is the McDonald function, $K_1(x) (x\to 0) \approx
\frac{1}{x}$,  so that for $t=0$ one obtains
\be
\varphi(0)=1.\label{a.2}\ee For $t>0$ the integration in
(\ref{a.1}) yields two different forms; e.g. for $t<1$,
\be
\varphi(t) =- \frac{3t^2}{(1-t^2)^{5/2}} \ln
\frac{1+\sqrt{1-t^2}}{t}+ \frac{1+2t^2}{(1-t^2)^2}\label{a.3}\ee
while for $t>1 $ one has instead,
\be
\varphi(t) =- \frac{3t^2}{(t^2-1)^{5/2}}\arctan
(\sqrt{t^2-1})+\frac{1+2t^2}{(1-t^2)^2}.\label{a.4}\ee For large
$t$ one has the following limiting behaviour, \be \varphi(t)
=\frac{2}{t^2} - \frac{3\pi}{2 t^3} + O(\frac{1}{t^4}).\label{a.5}
\ee

For small $t$ one obtains expanding the r.h.s. of (\ref{a.3})
\be
\varphi(t) =1 +t^2 (4-3\ln \frac{2}{t}) + t^4 (\frac74 -
\frac{15}{2} \ln \frac{2}{t} )+ O(t^6). \label{a.6} \ee
 Some numerical
values are useful in applications. $$\varphi(0.175) \cong 0.88,
~~\varphi (1.7) \cong 0.234, \varphi (5) \cong 0.052$$

\end{document}